\RequirePackage{fixltx2e}
\documentclass[aps,prb,reprint,floatfix, superscriptaddress]{revtex4-1}
\usepackage{latexsym}
\usepackage{amsmath}
\usepackage{amsfonts}
\usepackage{amssymb}
\usepackage{amsthm}
\usepackage{physics}
\usepackage{qcircuit}
\usepackage{graphicx}
\usepackage{dcolumn}
\usepackage{caption}
\usepackage{bm}
\usepackage{algorithm}
\usepackage{algpseudocode}
\usepackage{xcolor}
\usepackage{tabularx}
\usepackage{hyperref}
\AtBeginDocument{\usepackage{booktabs}} 

\newcommand{\eq}[1]{Eq.~(\ref{#1})} %
\newcommand{\fig}[1]{Fig.~\ref{#1}} %
\newcommand{\eqs}[1]{Eqs.~(\ref{#1})} %
\def\be{\begin{equation}} %
\def\ee{\end{equation}} %
\def\bea{\begin{eqnarray}} %
\def\eea{\end{eqnarray}} %
\newcommand{\kh}{\hat \kappa}

\newcommand{\hsz}{\hat S_z}
\newcommand{\HSS}{\hat S^2}
\newcommand{\LA}[1]{\mathfrak{#1}}
\newcommand{\HG}{\hat G}
\newcommand{\CR}[1]{\hat a^\dagger_{#1}}
\newcommand{\AN}[1]{\hat a_{#1}}
\newcommand{\BC}[1]{\color{black}{#1}}
\newcommand{\PTV}{\partial_\tau \hat V}
\begin{document}
\title{Analytic gradients in variational quantum algorithms: Algebraic extensions of the parameter-shift rule 
to general unitary transformations} 
  
  \author{Artur F. Izmaylov}
\email{artur.izmaylov@utoronto.ca}
\affiliation{Department of Physical and Environmental Sciences,
  University of Toronto Scarborough, Toronto, Ontario, M1C 1A4,
  Canada}
\affiliation{Chemical Physics Theory Group, Department of Chemistry,
  University of Toronto, Toronto, Ontario, M5S 3H6, Canada}
  
 \author{Robert A. Lang}
\affiliation{Department of Physical and Environmental Sciences,
  University of Toronto Scarborough, Toronto, Ontario, M1C 1A4,
  Canada}
\affiliation{Chemical Physics Theory Group, Department of Chemistry,
  University of Toronto, Toronto, Ontario, M5S 3H6, Canada}
  
 \author{Tzu-Ching Yen}
\affiliation{Chemical Physics Theory Group, Department of Chemistry,
  University of Toronto, Toronto, Ontario, M5S 3H6, Canada}

\begin{abstract}
Optimization of unitary transformations in Variational Quantum Algorithms   
benefits highly from efficient evaluation of cost function gradients with respect to amplitudes of unitary generators. 
We propose several extensions of the parametric-shift-rule to formulating these gradients as linear combinations of 
expectation values for generators with general eigen-spectrum (i.e. with more than two eigenvalues). 
Our approaches are exact and do not use any auxiliary qubits, instead they rely on a generator 
eigen-spectrum analysis. Two main directions in the parametric-shift-rule extensions are 1) polynomial expansion of the 
exponential unitary operator based on a limited number of different eigenvalues in the generator and 2) decomposition of the generator
as a linear combination of low-eigenvalue operators (e.g. operators with only 2 or 3 eigenvalues). These 
techniques have a range of scalings for the number of needed expectation values with the number of generator 
eigenvalues from quadratic (for polynomial expansion) to linear and even $\log_2$ (for generator decompositions).   
{\BC This allowed us to propose efficient differentiation schemes superior to previous approaches 
for commonly used 2-qubit transformations (e.g. match-gates, 
transmon and fSim gates) and $\hat S^2$-conserving fermionic operators for the variational quantum eigensolver. } 
\end{abstract}

\maketitle

\date{\today}

\section{Introduction}

Variational Quantum Algorithms (VQA) currently provide the main route to employing noisy intermediate-scale quantum hardware 
without error correction to solve classically difficult optimization problems in quantum chemistry,\cite{Peruzzo2014,doi:10.1021/acs.chemrev.8b00803,RevModPhys.92.015003} information compression,\cite{Romero:2017ci} machine learning,\cite{PhysRevA.101.032308,Mitarai:2018ka,Benedetti:2019dh} and number factorization.\cite{Anschuetz:2018vb}  
The mathematical formulation of VQA involves a cost function defined as follows
\begin{align} \label{eq:E}
E(\boldsymbol{\tau}) =  \bra{\bar 0} \hat U^\dagger (\boldsymbol{\tau}) \hat H \hat U (\boldsymbol{\tau})\ket{\bar 0},
\end{align}
where $\hat H$ is some hermitian $N$-qubit operator (e.g. the quantum system Hamiltonian for quantum chemistry applications), 
$\hat U (\boldsymbol{\tau})$ is a unitary transformation encoded on a quantum computer as a circuit operating on 
the initial state of $N$ qubits $\ket{\bar 0} \equiv \ket{0}^{\otimes N}$. 
To avoid deep circuits, $E(\boldsymbol{\tau})$ is optimized with respect to $\boldsymbol{\tau}$ components 
using a hybrid quantum-classical iterative process: 1) every set of $\boldsymbol{\tau}$ parameters is implemented on 
a quantum computer to measure value of $E(\boldsymbol{\tau})$, 2) results of quantum measurements are passed to 
a classical computer to suggest a next set of $\boldsymbol{\tau}$ parameters.   

Naturally, this hybrid scheme becomes more efficient if a quantum computer can provide gradients of 
$E(\boldsymbol{\tau})$ with respect to $\boldsymbol{\tau}$ components. 
In molecular problems, analytical gradients with respect to circuit parameters are not only useful for variational energy 
optimization, but also in the calculation of analytical nuclear energy gradients and non-adiabatic couplings in 
variational quantum eigensolver extensions for excited states.\cite{arimitsu2021analytic,yalouz2021analytical}
Usual parametrizations of unitary transformations are organized as products of exponential functions of some hermitian 
generators $\{\HG_k\}$,  
\bea
\hat U (\boldsymbol{\tau}) = \prod_k \exp(i\tau_k \hat G_k).
\eea 
The choice of efficient generators is generally a challenging problem whose solution often relies on heuristics of a concrete 
field (e.g. in quantum chemistry there is a large variety of techniques developed recently \cite{Romero2018,Ryabinkin2018,Ryabinkin2019b,qAdapt,Gard:2019vd,Lee:2019/jctc/311,Sokolov2019,Dallaire_Demers2019,PRXQuantum.2.020337,LangILC:2021}). 
Due to general non-commutativity of generators, $\tau_k$ gradients can be written as
\bea\notag
 \frac{\partial E}{\partial \tau_k}&=& \frac{\partial}{\partial \tau_k}\bra{\bar 0} \hat U_1^\dagger e^{-i\tau_k \hat G_k} \hat U_2^\dagger \hat H \hat U_2 e^{i\tau_k \hat G_k}\hat U_1\ket{\bar 0} \\\label{eq:epd}
 &=& i \bra{\bar 0} \hat U_1^\dagger e^{-i\tau_k \hat G_k} [\hat U_2^\dagger \hat H \hat U_2,\hat G_k] e^{i\tau_k \hat G_k}\hat U_1\ket{\bar 0},
\eea
where $\hat U_{1,2}$ are $\hat U$ parts on the left and right sides of the $\hat G_k$ exponent. Evaluating the gradient as the 
expectation value in \eq{eq:epd} requires extra efforts to accommodate for non-symmetric distribution of unitary 
transformations around $\hat H$ (considering the simplest case when $\hat G_k$ is also unitary). This treatment requires 
introducing an auxiliary qubit and controlled unitaries in the circuit, which enhance depth of the circuit.\cite{Schuld:2018gx} 

It was found that in case when $\hat G_k$ has only two eigenvalues symmetrically distributed, $\{\pm \lambda\}$, the so-called
parametric-shift-rule (PSR) is applicable to \eq {eq:epd}\cite{PhysRevLett.118.150503,Schuld:2018gx} 
\bea\notag
 \frac{\partial E}{\partial \tau_k}&=& \lambda\Big(\bra{\bar 0} \hat U_1^\dagger e^{-i(\tau_k+s) \hat G_k} \hat U_2^\dagger \hat H \hat U_2 e^{i(\tau_k+s) \hat G_k}\hat U_1\ket{\bar 0} \\ \label{eq:PSR}
 &&- \bra{\bar 0} \hat U_1^\dagger e^{-i(\tau_k-s) \hat G_k} \hat U_2^\dagger \hat H \hat U_2 e^{i(\tau_k-s) \hat G_k}\hat U_1\ket{\bar 0}\Big),~
\eea
where $s = \pi/(4\lambda)$. Clearly, this approach allows one to evaluate the expectation values in \eq{eq:PSR} using 
the same circuit as for $E(\boldsymbol{\tau})$ with only minor modifications of $\boldsymbol{\tau}$ parameters. 

A natural question is how to extend the PSR to a general unitary transformation containing generators with more than 2 eigenvalues. 
The algebraic form of these extensions is set to be a linear combination of expectation values of \eq{eq:PSR}. 
Such extensions are motivated by active developments in hardware (2-qubit gates\cite{Chow:2011ir,Foxen:2020jj}, 
generally have 4 eigenvalues) and theory related to specific problems (new generators for solving quantum chemistry problems,
e.g. spin-adapted fermionic rotations). 
Also, these extensions will allow one to reduce the number of optimized parameters if more complex generators can be considered. 


Generators with more than two eigenvalues can naturally be decomposed to generators with two eigenvalues 
for which the PSR can be applied individually, as suggested by Crooks.\cite{Crooks:2019tj} 
Few approaches to such decompositions were considered for some standard 2-qubit gates.\cite{Crooks:2019tj} 
But there was no attempt to systematically address the minimization of the number of terms in such decompositions 
or their extensions beyond 2-qubit operators. 

A naive application of the Ref.~\citenum{Crooks:2019tj} decomposition scheme to generators of the unitary coupled cluster (UCC) approach can lead to exponential growth of the number of terms (Pauli products) with two eigenvalues.\cite{Kottmann:2020js} Using a specifically tailored decomposition of UCC generators (fermionic-shift rule) Ref.~\citenum{Kottmann:2020js} was able to address the exponential growth of expectation values needed for gradient evaluation in the Pauli product decomposition of UCC generators. 
 
 
Here we demonstrate how to do the generator decomposition systematically and 
how to avoid cases of exponential increase of terms in such decompositions.  
We provide three generalizations of the PSR based on somewhat different algebraic ideas whose main unifying  
theme is consideration of the generator eigen-spectrum. In the first approach, we use the fact that the exponential function of the  
generator with $K$ eigenvalues can be presented as a $K-1$ degree polynomial. 
This allows us to extend the PSR by using a larger number of expectation values in a linear combination 
to cancel all higher powers of the generator.    
In the second approach, we decompose the generator into a sum of commuting operators with a fewer number of 
unique eigenvalues. The third approach uses a decomposition over non-commutative operators with 
a low number of eigenvalues. While the first approach can be seen as a generalization of Ref.~\citenum{Kottmann:2020js}
ideas to generators beyond those found in UCC, the second and third approaches have no apparent 
connections with previous works on efficient evaluation of gradients.  

Note that even though there are multiple generalizations of the PSR to higher derivatives for amplitudes of generators with two eigenvalues\cite{hubregtsen2021singlecomponent,PhysRevA.103.012405} and stochastic techniques for gradients of an arbitrary generator,\cite{Banchi2021measuringanalytic} they will not be considered here, since our focus is on  deterministic expression for gradients involving expectation values of hermitian operators.

One problem that is related to using analytical gradients in variational algorithms is the problem of barren plateaus (exponentially vanishing gradients).\cite{McClean_2018,Cerezo_2021,PhysRevLett.126.140502} The theoretical developments discussed in this paper do not consider this problem, since it is related to specifics of the variational problem and a choice of generators.

\section{Theory}

\subsection{Polynomial expansion}

Energy partial derivatives (\eq{eq:epd}) can be rewritten as
\bea\label{eq:grad}
 \frac{\partial E}{\partial \tau}= i\langle e^{-i\tau \hat G} \hat H_2 \hat G e^{i\tau \hat G}\rangle - i\langle \hat G e^{-i\tau \hat G} \hat H_2 e^{i\tau \hat G}\rangle,
\eea
where we removed $k$ subscript for simplicity and introduced short notation: $\langle ... \rangle = \bra{\bar 0} \hat U_1^\dagger ... \hat U_1\ket{\bar 0}$, 
$\hat H_2 = \hat U_2^\dagger \hat H \hat U_2$.  
Asymmetry of operators' placement around $\hat H$ and potential non-unitarity of $\hat G$ makes the obtained expectation values more 
challenging to measure.  
However, if $\hat G$ has a finite number of different eigenvalues, 
there is a way to rewrite this difference as a linear combination of terms measurable on a quantum computer without any modifications 
of the $E(\boldsymbol{\tau})$ measurement scheme
\bea\label{eq:exp}
 \frac{\partial E}{\partial \tau}&=& \sum_n C_n \langle e^{-i(\tau+\theta_n) \hat G} \hat H_2 e^{i(\tau+\theta_n) \hat G}\rangle,\quad
\eea
where $\theta_n$ and $C_n$ are coefficients to be defined. Details on obtaining this generalization of the PSR are given in Appendix A. The key quantity that defines $\theta_n$ and $C_n$ is the number 
of different eigenvalues in $\hat G$, which will be denoted as $L$. $L$ defines a finite polynomial expression for the exponential operator
\bea\label{eq:poly}
e^{i\theta \hat G} = \sum_{n=0}^{L-1} a_n(\theta) (i\hat G)^n, 
\eea
where $a_n(\theta)$'s are constants obtained by solving a linear system of equations, and 
$C_n$'s are evaluated from another system of linear equations using $a_n(\theta)$'s with fixed $\theta_n$'s 
(see Appendix A for further details). To illustrate the process in the simplest case 
of $L=2$, where $\hat G$'s eigenvalues are $\pm 1$, hence $\hat G^2 = \hat 1$ and 
\bea\label{eq:invol}
e^{i\theta \hat G} =  a_0(\theta) \hat 1 + a_1(\theta) (i\hat G), 
\eea 
here, $a_0(\theta) = \cos(\theta)$ and $a_1(\theta) = \sin(\theta)$. To obtain the energy derivative we need 
only two terms in \eq{eq:exp}
\bea\notag
 \frac{\partial E}{\partial \tau}&=& \frac{1}{\sin(2\theta)}\Bigg[\langle e^{-i(\tau+\theta) \hat G} \hat H_2 e^{i(\tau+\theta) \hat G}\rangle \\ \label{eq:exp1}
 &&-  \langle e^{-i(\tau-\theta) \hat G} \hat H_2 e^{i(\tau-\theta) \hat G}\rangle\Bigg].
\eea
For example, $\hat G$ satisfying the described conditions can be any tensor product of Pauli operators for different qubits.

For $L=3$ with symmetric spectrum $\{0,\pm 1\}$, Appendix A shows that 4 expectation values 
are enough to obtain the analytic gradient. It was shown recently that all 
fermionic operators $\hat G = \CR{p}...\CR{q}\AN{r}...\AN{s} - \CR{s}...\CR{r}\AN{q}...\AN{p}$
used in the UCC method have this spectrum.\cite{Kottmann:2020js} Techniques developed in Ref.~\citenum{Kottmann:2020js} also 
found the gradient expressions requiring 4 expectation values for such operators, and were able to reduce it to only 2 expectation values 
for real unitary transformations acting on real wave-functions. The polynomial expansion can be seen as a generalization 
of  Ref.~\citenum{Kottmann:2020js} ideas to generators beyond the fermionic rotations used in the UCC method.  

The polynomial expansion for general $\hat G$ with $L$ eigenvalues will produce the gradient expression with the
number of expectation values that scales as $\sim L^2$. If there are some relations between different eigenvalues,
they can be used to reduce the number of expectation values by exploiting freedom in the choice of $\theta_n$ and $C_n$ 
parameters (see Appendix A for more details). 

\subsection{Generator decompositions}

To address the $\sim L^2$ scaling of the number of expectation values in the polynomial expansion approach, 
instead of \eq{eq:exp} we will use the following alternative
\bea
 \frac{\partial E}{\partial \tau} = \sum_n C_n \langle e^{-i\theta_n \hat O_n} e^{-i\tau\hat G} \hat H_2 e^{i\tau \hat G} e^{i\theta_n\hat O_n}\rangle,\quad\label{eq:exp2}
\eea 
where we introduced new operators $\{\hat O_n\}$. 
$\hat O_n$'s are required to have only a few eigenvalues (e.g. 2 or 3) and to sum to $\hat G$
\bea\label{eq:GXn}
\hat G = \sum_{n=1}^{K} d_n\hat O_n, 
\eea
where $d_n$ are real coefficients.
\paragraph{Involutory example:}
To illustrate how $\{\hat O_n\}$-decomposition can be used in the gradient evaluation, let us assume 
that $\hat O_n$'s have only two eigenvalues $\pm 1$.  To define $C_n$ and 
$\theta_n$ let us consider the following pairs
\bea \notag
&&\langle e^{-i \theta_n \hat O_n} e^{-i\tau\hat G} \hat H_2 e^{i\tau \hat G} e^{i\theta_n \hat O_n} \rangle 
-  \langle e^{i \theta_n \hat O_n} e^{-i\tau\hat G} \hat H_2 \\ \notag
&&\times e^{i\tau \hat G} e^{-i\theta_n \hat O_n} \rangle 
= i\sin(2\theta_n)\Bigg[\langle e^{-i\tau \hat G} \hat H_2 e^{i\tau \hat G} \hat O_n \rangle  \\
&-&  \langle \hat O_n e^{-i\tau \hat G} \hat H_2 e^{i\tau \hat G}\rangle\Bigg].
\eea
Here, we used the involutory property of $\{\hat O_n\}$ to convert their exponents according to \eq{eq:invol}.
This consideration shows that to obtain the energy derivative via expansion in \eq{eq:exp2}
we should select $\pm\theta_n$ pairs with coefficient $C_{\pm n} = d_n/\sin(\pm 2\theta_n)$. 
The number of the expectation values in \eq{eq:exp2} is $2K$. 
Unfortunately, $K$ depends not only on the number of $\hat G$ eigenvalues but also on their 
distribution and degeneracies (or multiplicities).
However, it is easy to formulate the best case scenario where $K= \log_2(L)$, here adding $K$ 
$\hat O_n$ operators produces $\hat G$ whose spectrum has $2^K$ eigenvalues $\{\lambda_j\}$
\bea
\lambda_j = \sum_{n=1}^K d_n b_{nj}, ~ b_{nj} = \{\pm 1\}. 
\eea
Starting with some $\hat G$, it is not necessary that its eigenvalues will be encoded so efficiently 
with involutory operators $\hat O_n$'s, yet this best case scenario shows great potential for the 
decomposition approach. 

\paragraph{Efficient generator decompositions:} $\hat O_n$ operators 
optimal for the generator decomposition depend on the spectrum of $\hat G$. 
We assume that $\hat G$ can be written in terms of a few qubit or fermionic operators. The number 
of involved qubits or fermionic spin-orbitals should not exceed the limit when the 
dimensionality of a faithful representation for involved operators becomes too large to 
do matrix algebra on a classical computer.

The mathematical basis for representing $\hat G$ as matrix $\mathbf{G}$ is that qubit or 
fermionic operators expressing $\hat G$ can be considered as basis elements of a Lie algebra. 
Using a faithful representation of this Lie algebra one can work with corresponding 
matrices instead of operators. $\mathbf{G}$ can be diagonalized  
$\mathbf{G} = \mathbf{V}^\dagger \mathbf{D}\mathbf{V}$ to obtain the guidance on choice of 
optimal $\hat O_n$'s. To minimize the number of $\hat O_n$ operators, 
one would build them from decomposition $\mathbf{D} = \sum_n \mathbf{D}_n$, where 
$\mathbf{D}_n$ are diagonal matrices with a few (2-3) different eigenvalues. 
Then, $\hat O_n$ is obtained via the inverse representation map of 
$\mathbf{O}_n = \mathbf{V}^\dagger \mathbf{D}_n\mathbf{V}$. 
The caveat is that even the decomposition of the diagonal matrix $\mathbf{D}$ can be done in various ways 
differing in the number of necessary $\mathbf{O}_n$'s. 
A simple example illustrating various possibilities is
\bea\notag
\begin{pmatrix} 
3 & 0 & 0 & 0 \\
0 & -3 & 0 &  0 \\
0 & 0 & -1 & 0 \\
0 & 0 & 0 & 1
\end{pmatrix} &=& \begin{pmatrix} 
3 & 0 & 0 & 0 \\
0 & -3 & 0 &  0 \\
0 & 0 & 0 & 0 \\
0 & 0 & 0 & 0
\end{pmatrix}+
\begin{pmatrix} 
0 & 0 & 0 & 0 \\
0 & 0 & 0 &  0 \\
0 & 0 & -1 & 0 \\
0 & 0 & 0 & 1
\end{pmatrix} \\\notag
&=& \begin{pmatrix} 
2 & 0 & 0 & 0 \\
0 & -2 & 0 &  0 \\
0 & 0 & -2 & 0 \\
0 & 0 & 0 & 2
\end{pmatrix}+
\begin{pmatrix} 
1 & 0 & 0 & 0 \\
0 & -1 & 0 &  0 \\
0 & 0 & 1 & 0 \\
0 & 0 & 0 & -1
\end{pmatrix} \\\label{eq:ve}
&=& 3(\mathbf{P}_1-\mathbf{P}_2) + (\mathbf{P}_4-\mathbf{P}_3),
\eea
where $\mathbf{P}_j$ is a 4 by 4 matrix with 1 on $(j,j)^{\rm th}$ element and zeroes everywhere else.
In this example, the most optimal choice is the second expansion, 2 operators with 2 symmetric 
eigenvalues each. It also shows that even though the eigen-subspace projector expansion (last in \eq{eq:ve}) 
is the most straightforward, it is not necessarily the most optimal.   

Using the form of the $\mathbf{D}_n$ matrix one can optimize the number and the form of $\hat O_n$
operators for a particular generator. The result of this optimization is not explicitly representable in some closed form
for an arbitrary generator. Instead, here we provide several heuristics that can generate shorter 
expansions than those from the eigen-subspace projector expansion, the latter 
can always be used as a conservative option.  

\paragraph{Commutative Cartan sub-algebra decomposition:}
The basis of our algebraic heuristics is a Cartan sub-algebra (CSA) decomposition for $\hat G$.\cite{CSA2020}
This decomposition can be done for an element of any compact Lie algebra. Here 
we will use it for $\hat G$ realized as an element of the $N$-qubit operator algebra, $\LA{su}(2^N)$, 
\bea
\hat G = \sum_n C_n \hat P_n, ~\hat P_n = \prod_{j=1}^{N} \hat \sigma_j, 
\eea
where $C_n$ are coefficients, and $\hat \sigma_j = \{\hat x_j, \hat y_j, \hat z_j, \hat {1\!\! 1}_j\}$ is one of the 
Pauli operator or identity for the $j^{\rm th}$ qubit. $\LA{su}(2^N)$ contains $4^N-1$ generators $\hat P_n$ due 
to the exclusion of the tensor product of $N$ identity operators. The largest abelian (or Cartan) sub-algebra 
in $\LA{su}(2^N)$ that we will involve in the decomposition is a set of $\hat P_n$'s that contain 
only $\hat z_j$ operators, denotes as $\hat Z_n$'s. $\hat Z_n$'s have only 2 distinct eigenvalues, $\pm 1$, which is convenient for our decomposition.
The CSA decomposition of $\hat G$ is    
\bea \label{eq:G}
\hat G = \hat V^\dagger \left(\sum_{n=1}^K c_n \hat Z_n \right)\hat V,
\eea
where $c_n$ are coefficients, and $\hat V$ is a unitary transformation 
\bea
\hat V = \prod_k e^{i\tau_k \hat P_k},
\eea
here, $\tau_k$ are real amplitudes, and $\hat P_k$'s are all Pauli products that are not in the CSA.
Clearly, each term in the sum of \eq{eq:G} has eigenvalues 
$\pm c_n$, therefore we can choose each $\hat O_n = c_n \hat V^\dagger \hat Z_n \hat V$. 

The CSA decomposition in \eq{eq:G} can be done by expanding the left- and right-hand sides of 
 \eq{eq:G} in a basis of $\LA{su}(2^N)$ Lie algebra of $\hat P_n$'s 
 to find coefficients $c_n$ and amplitudes $\tau_k$ for $\hat V$. This decomposition is unique 
 in terms of the number of $\hat Z_n$ terms, which suits our purpose to obtain the number 
 of two-eigenvalue operators in the $\hat G$ decomposition.  
 
Since all $\hat O_n$ operators commute, one can rewrite the gradient expression 
as application of the PSR to each $\hat O_n$ operator in $\hat G$
\bea\notag
 \frac{\partial E}{\partial \tau} &=& \sum_n \frac{1}{\sin(2\theta_n)}\Bigg[
 \langle e^{-i(\tau\hat G+\theta_n \hat O_n)} \hat H_2 e^{i(\tau \hat G+\theta_n\hat O_n)}\rangle \\
 &-& \langle e^{-i(\tau\hat G-\theta_n \hat O_n)} \hat H_2 e^{i(\tau \hat G-\theta_n\hat O_n)}\rangle\Bigg]. \label{eq:exp3}
\eea 
The involved unitary transformations can be rewritten as 
\bea
e^{\pm i(\tau\hat G \pm \theta_n \hat O_n)} = \hat V^\dagger \prod_{m=1}^K e^{\pm i c_m (\tau \pm \delta_{nm}\theta_n) \hat Z_m }\hat V.
\eea
This form is convenient for implementation of these operators as a circuit. 

\paragraph{Non-commutative Cartan sub-algebra decomposition:}
An alternative representation of $\hat G$ is a sum of non-commuting two-eigenvalue operators  
\bea \label{eq:G2}
\hat G = \sum_{n=1}^{K'} c_n \hat V_n^\dagger \hat Z_n \hat V_n,
\eea
here, $\hat V_n$'s are defined in the same way as $\hat V$.  
This decomposition defines $\hat O_n = c_n \hat V_n^\dagger \hat Z_n \hat V_n$, and due to differences in 
$\hat V_n$'s, different $\hat O_n$'s do not necessarily commute. 
The main advantage of the non-commutative decomposition is that 
it uses not only coefficients $c_n$ for reproducing the spectrum of $\hat G$ but 
also some parameters in $\hat V_n$'s, $\lambda_j = \lambda_j (\{V_{n}\}_{n=1}^{K'}, \{c_{n}\}_{n=1}^{K'})$.
This dependence provides an opportunity for the non-commutative decomposition 
to represent $\hat G$ with a lower number of terms $K'<K$ (cf. \eq{eq:G2} and \eq{eq:G}). 

To construct the non-commutative decomposition we fix the number of terms $K'$ to 
values lower than $K$ in \eq{eq:G} and minimize the difference between the left- and 
right-hand sides of \eq{eq:G2} using $c_n$ and $\tau_k^{(n)}$ (amplitudes of $\hat V_n$). 
The choice of $\hat Z_n$ in \eq{eq:G2} is insignificant because $\hat V_n$ can always transform one 
CSA operator into another. 

Non-commutativity of $\hat O_n$ operators does not preclude 
use of the shift-rule to each $\hat O_n$ operator to obtain components of the derivative for 
the $\hat G$ amplitude
\bea\notag
 \frac{\partial E}{\partial \tau} &=& \sum_n \frac{1}{\sin(2\theta_n)}\Bigg[
 \langle e^{-i\theta_n \hat O_n} e^{-i\tau\hat G}  \hat H_2 e^{i\tau\hat G} e^{i\theta_n\hat O_n}\rangle \\
 &-& \langle e^{i\theta_n \hat O_n} e^{-i\tau\hat G}  \hat H_2 e^{i\tau\hat G} e^{-i\theta_n\hat O_n}\rangle\Bigg]. \label{eq:exp4}
\eea 
To measure such expectation values there is overhead related to non-compatibility of eigenstates for 
individual $\hat O_n$ and $\hat G$. Thus, one needs to explore for each class of $\hat G$ operators, whether 
the potential reduction in the number of terms in \eq{eq:G2} is not diminished by a possible higher circuit depth. 


\section{Applications}

We will consider application of the generator decompositions for gradient evaluations of several classes of challenging operators: 1) 2-qubit generators, 2) 3-qubit generators, and 3) generators of $\hat S^2$-conserving fermionic rotations. Our choice was motivated not only by inapplicability of the PSR for these generators due to the multitude of eigenvalues but also because advantages of all three decomposition techniques can be illustrated on them. To compare results of the proposed decompositions with a previous general scheme from Ref.~\citenum{Schuld:2018gx}, we start this section with reviewing the latter.   

\subsection{Gradients via linear combination of unitaries}

Denoting $\hat V = e^{i\tau \hat G}$ and $\PTV=i \hat G e^{i\tau \hat G}$, one can rewrite \eq{eq:grad} as 
\bea
\frac{\partial E}{\partial \tau}= \langle \hat V^\dagger \hat H_2 \PTV \rangle + \langle \PTV^\dagger \hat H_2 \hat V\rangle.
\eea
To use a measurement scheme introduced in Ref.~\citenum{Schuld:2018gx}, $\PTV$ needs to be decomposed as a 
linear combination of unitaries (LCU). 
Since $e^{i\tau \hat G}$ is already a unitary operation, the decomposition is only needed for the $i\hat G$ part
\bea\label{eq:PV}
\PTV = \sum_{k=1}^{K_u} c_k \hat W_k e^{i\tau\hat G},
\eea  
where $i\hat G$ is decomposed in linear combinations of unitaries $\hat W_k$ with coefficients $c_k$. A typical choice of 
$\hat W_k$'s is a set of Pauli products $i\hat P_k$ comprising $\hat G$. Also, one can use the commutative CSA decomposition 
\eq{eq:G} to obtain potentially more compact set of $\hat W_k$'s. Note though that the decomposition in 
\eq{eq:PV} is less flexible than the one in \eq{eq:GXn} because the latter does not require unitarity of $\hat O_n$ operators. 

\begin{figure}[h!]
  \includegraphics[width=0.9\columnwidth]{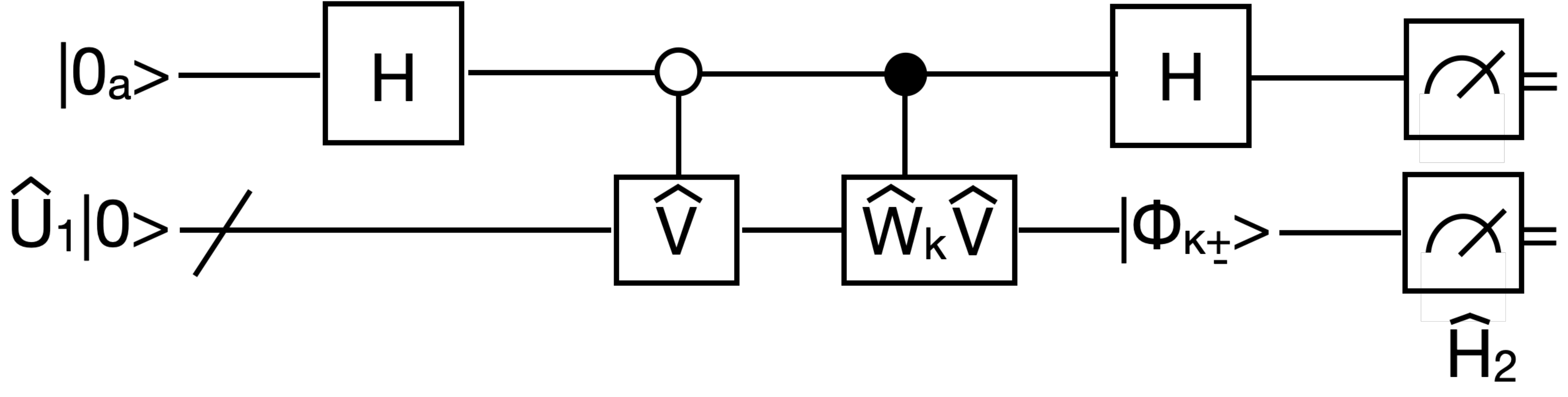}
  \caption{Measurement scheme for terms in \eq{eq:mLCU}. States $\ket{\Phi_{k\pm}} = (\hat V\pm \hat W_k\hat V)\hat U_1\ket{0}/(2\sqrt{p_{\pm}})$ are obtained after measurement of the ancilla qubit, 
  where $p_{\pm} = \bra{0}\hat U_1^\dagger (\hat V\pm\hat W_k\hat V)^\dagger (\hat V\pm\hat W_k\hat V)\hat U_1\ket{0}/4$ is the probability for the ancilla qubit results to be $\pm 1$ respectively. $\rm H$ is the Hadamard gate.}
  \label{fig:LCUc}
\end{figure}

The LCU decomposition for $\PTV$ allows one to rewrite  
\bea
\frac{\partial E}{\partial \tau}&=& \sum_{k=1}^{K_u} c_k [\langle \hat V^\dagger \hat H_2 \hat W_k \hat V\rangle + c.c.]\\ \notag
&=& \frac{1}{2}\sum_{k=1}^{K_u} c_k [\langle \hat V^\dagger (1+\hat W_k)^\dagger\hat H_2 (1+\hat W_k) \hat V\rangle  \\\label{eq:mLCU}
&-& \langle \hat V^\dagger (1-\hat W_k)^\dagger \hat H_2 (1-\hat W_k) \hat V\rangle].
\eea
Each term in square brackets of \eq{eq:mLCU} can be obtained via quantum measurement using the circuit depicted in \fig{fig:LCUc}.
Extra features required for the circuit on \fig{fig:LCUc} are one ancilla qubit and controlled versions of unitaries. These features 
are not needed for a regular circuit measuring expectation values in \eqs{eq:exp} and \eqref{eq:exp2}. 
In what follows we will only compare the number of expectation values needed to be measured in the LCU scheme and the 
proposed approaches. If this number is lower or the same in the new schemes, absence of the ancilla qubit and 
controlled unitaries in the new schemes make these approaches more advantageous than the LCU scheme.

\subsection{2-qubit generators}

Any 2-qubit generator has not more than 4 different eigenvalues, and thus, the eigenvalue decomposition scheme 
will need 8 expectation values for a gradient evaluation. 
The CSA decomposition (\eq{eq:G}) of any 2-qubit generator results in at most 3 $\hat Z_n$'s ($\hat z_1, \hat z_2, \hat z_1 \hat z_2$), which leads 
to not more than 6 expectation values for each gradient. In all considered 2-qubit gates, 
commuting and non-commuting CSA decompositions provided the same number of $\hat O_n$'s.  
Since all $\hat O_n$'s provided by CSA decompositions are unitary, they can be used in the LCU 
scheme. Therefore, the number of terms required for measurements in the LCU and new schemes are the 
same, but the LCU scheme will require an extra qubit and controlled versions of unitary operations.


{\it Transmon gates:} These gates are generated by\cite{Chow:2011ir} 
\bea
  \hat G &=& \hat x_1 - b\hat z_1\hat x_2 + c\hat x_2.
\eea
Applying $\hat W(\tau) = \exp(i\tau\hat y_1 \hat x_2)$ to each term of $\hat G$
\bea
  \hat W^\dagger(\tau) \hat x_1 \hat W(\tau) &=& \cos(2\tau) \hat x_1 - \sin(2\tau) \hat z_1 \hat x_2 \\
  \hat W^\dagger(\tau) \hat z_1\hat x_2 \hat W(\tau) &=& \cos(2\tau) \hat z_1 \hat x_2 + \sin(2\tau) \hat x_1\\
  \hat W^\dagger(\tau) \hat x_2 \hat W(\tau)  &=& \hat x_2 
\eea 
one can choose $\tau_0$ so that $\cos(2\tau_0) = 1/\sqrt{1+b^2}$ and $\sin(2\tau_0) = b/\sqrt{1+b^2}$, 
then $\hat G$ can be represented as 
\bea
\hat G = \hat W^\dagger(\tau_0) [\sqrt{1+b^2} \hat x_1 +c \hat x_2] \hat W(\tau_0). 
\eea
To arrive at the form of \eq{eq:G}, $\hat V$ needs to be defined as 
\bea
\hat V = e^{i\frac{\pi}{4}(\hat y_1+\hat y_2)} \hat W(\tau_0)
\eea
then $\hat O_1 = \sqrt{1+b^2} \hat V^\dagger \hat z_1 \hat V$ and $\hat O_2= c\hat V^\dagger \hat z_2 \hat V$.
This decomposition allows one to evaluate the gradient using only 4 expectation values. 

{\it Match-gates:} Generators of these gates are linear combinations of the following operators\cite{Dallaire_Demers2019,Jozsa:2008cz} 
\bea
  \{\hat x_1 \hat x_2,\ \hat y_1\hat y_2,\ \hat x_1 \hat y_2,\ \hat y_1 \hat x_2,\ \hat z_1,\ \hat z_2 \}.
\eea 
This set forms a sub-algebra of $\LA{su}(4)$ that is a direct sum of two $\LA{su}(2)$ algebras 
\bea
  \mathcal{A}_1 &=& \left\{
    \frac{\hat z_1 + \hat z_2}{2}, 
    \frac{\hat x_1\hat y_2 + \hat y_1\hat x_2}{2}, 
    \frac{\hat x_1\hat x_2 - \hat y_1\hat y_2}{2} 
  \right\} \\ 
  \mathcal{A}_2 &=& \left\{
    \frac{\hat z_1 - \hat z_2}{2}, 
    \frac{\hat y_1\hat x_2 - \hat x_1\hat y_2}{2}, 
    \frac{\hat x_1\hat x_2 + \hat y_1\hat y_2}{2}
   \right\}.
\eea 
Each $\LA{su}(2)$ has only one Cartan element. 
The CSA decomposition of any match-gate generator provides two $\hat O_n$'s, which are results of  
conjugation of two CSA elements, $\hat z_1$ and $\hat z_2$, with unitaries ($\hat V$'s) from the two $SU(2)$ groups 
corresponding to the $\LA{su}(2)$ algebras. 

{\it fSim gates:} The fSim gate generator\cite{Arute_2019,Foxen:2020jj} is 
\bea
  \hat G_{\rm fSim} = \frac{\theta}{2} (\hat x_1\hat x_2 + \hat y_1\hat y_2) 
  + \frac{\phi}{4} (1 - \hat z_1) (1 - \hat z_2). 
\eea 
Its CSA decomposition results in 3 $\hat Z_n$, therefore to do gradients with respect to the overall amplitude 
$\tau$ in $\exp(i\tau \hat G_{\rm fSim})$ will require 6 expectation values. $\hat G_{\rm fSim}$ can be split into 
\bea
  \hat O_1 &=& \frac{\theta}{2}(\hat x_1\hat x_2 + \hat y_1\hat y_2), \\
  \hat O_2 &=& \frac{\phi}{4} (1 - \hat z_1) (1 - \hat z_2),
\eea 
which have 3 and 2 eigenvalues respectively, thus the $\theta$ and $\phi$ gradients of 
$\exp(i\hat G_{\rm fSim}(\theta,\phi))$ will require 4 and 2 expectation values. 


\subsection{3-qubit generators}

Considering the 2-qubit generators, we were not able to find a case where the non-commutative CSA decomposition 
had an advantage over the commutative CSA scheme. Here, we give an example of a 3-qubit transformation 
where this advantage is clear. Consider a generator 
\bea
  \hat G &=& \hat U^\dagger \hat z_1 \hat U + \hat z_2
\eea 
that requires only 2 $\hat O_n$'s using the non-commuting scheme. 
We choose a three-qubit unitary $\hat U = \exp(\hat A)$, where $\hat A$ is an anti-Hermitian operator with the following matrix representation: $A_{i,i} = 0$, $A_{i,j<i}=1$, and $A_{i,j>i}=-1$. 
The CSA decomposition of $\hat G$
\bea
\hat G &\approx& \hat V^\dagger (1.250 \hat z_1 \hat z_2 + 
0.045 \hat z_1 \hat z_2 \hat z_3 + 
0.014 \hat z_1 \hat z_3 \nonumber\\ 
&&+
0.658 \hat z_2 
-0.045 \hat z_2 \hat z_3 + 
0.014 \hat z_3) \hat V 
\eea indicates that there are at least 6 $\hat O_n$'s (12 expectation values) for the commutative decomposition 
scheme. 

\subsection{$\hat S^2$-conserving fermionic generators}

One of the approaches to construct a pool of generators for application of VQAs to solving the electronic 
structure problems is adding symmetry conserving conditions.\cite{Izmaylov:2020hb} 
Usual UCC single and double operators 
\bea
\kh_i^a &=& a_a^\dagger a_i - a_i^\dagger a_a\\
\kh_{ji}^{ab} &=&  a_a^\dagger a_b^\dagger a_i a_j - a_j^\dagger a_i^\dagger a_b a_a
\eea
conserve the number of electrons but generally not the electron spin. 
Unitary generators that commute with the electron spin operators, $\hsz$ and $\HSS$, can be obtained by 
anti-hermitization of singlet spherical tensor operators.\cite{Helgaker} 
A general spherical tensor operator, $\hat T^{S,M}$, is defined as 
\bea
[\hat S_{\pm}, \hat T^{S,M}] = \sqrt{S(S+1)-M(M\pm 1)} \hat T^{S,{M\pm 1}},
\eea
\bea
[ \hsz, \hat T^{S,M} ] = M \hat T^{S,M} ,
\eea 
where $S$ and $M$ are electron spin and its projection to the $z$-axis, respectively. 
Equation $\HSS = \hat S_{-}\hat S_{+} + \hsz(\hsz+1)$ can be used to show that any 
singlet spherical tensor operator, $\hat T^{0,0}$ will commute with $\hsz$ and $\HSS$. 

There are standard approaches for producing spherical tensor operators,\cite{Helgaker}
they involve very similar techniques to those used for generating spin-adapted 
configuration state functions.\cite{Paldus,Shavitt}
Individual single excitations are not $\hat T^{0,0}$ operators, therefore, one needs to group 
more than one excitation to obtain singlet operators
\bea \label{eq:ucc_singlet_s}
\hat T^{0,0}_{ia} = \kh_{i_\alpha}^{a_\alpha} +\kh_{i_\beta}^{a_\beta},
\eea
here and further $a_\alpha$($a_\beta$) and $i_\alpha$($i_\beta$) are the spin orbitals arising 
from the $\alpha$($\beta$) spin parts of the $a^{\rm{th}}$ and $i^{\rm{th}}$ spatial orbitals.
For double and higher excitations/de-excitations, the seniority number $\Omega$ 
(the number of unpaired electrons created by the operator) correlates well with the number of 
 individual excitation/de-excitation pairs in construction of singlet operators
\bea \label{eq:ucc_singlet_dsen0}
\Omega=0: &\quad& \hat T^{0,0}_{iiaa} = \kh_{i_{\alpha} i_{\beta}}^{a_{\alpha} a_{\beta}},\\
\label{eq:ucc_singlet_dsen2}
\Omega=2: &\quad& \hat T^{0,0}_{iiab} = \kh_{i_{\alpha}i_{\beta}}^{a_{\alpha}b_{\beta}} + \kh_{i_{\alpha}i_{\beta}}^{a_{\beta} b_{\alpha}}, \\
\Omega=2: &\quad& \hat T^{0,0}_{ijaa} = \kh_{i_{\alpha}j_{\beta}}^{a_{\alpha}a_{\beta}} + \kh_{i_{\beta}j_{\alpha}}^{a_{\alpha}a_{\beta}}, \\
\label{eq:ucc_singlet_dsen4}
\Omega=4: &\quad& \hat T^{0,0}_{ijab} = \sum_{s, \bar{s} \in \{\alpha, \beta \}} \kh_{i_{s}j_{\bar{s}}}^{a_{\bar{s}}b_{s}}.
\eea
Generators in \eqs{eq:ucc_singlet_s}-\eqref{eq:ucc_singlet_dsen4} are required for spin-conserving UCC singles and doubles ansatz. Note that these generators can be used to add electronic correlation to an initial state of any electron spin symmetry (not necessarily closed-shell singlet) without altering the spin state.

The spectrum of the spin-conserving generators are reported in Table \ref{tab:singlet_spectra}. It is important to note that the zero eigenvalue 
has much higher multiplicity than the non-zero eigenvalues for the single and double spherical tensor operators. Due to large differences between
multiplicities of different eigenvalues in the singlet operators' spectra, their decomposition following \eq{eq:G} was found to be inefficient in $K$.   
We found that it usually takes a lot of $\{\pm 1 \}$-eigenvalued operators to create large variations in eigenvalues' multiplicities. Furthermore, due to parity symmetry of the spectra, it is natural to introduce alternative $\hat O_n$'s in \eq{eq:exp2} which have 3 eigenvalues $\{0, \pm \lambda_n \}$.
\begin{table}[h]%
\setlength\tabcolsep{0pt}
\caption{The eigenvalues and number of Pauli products for the singlet single and double fermionic operators. 
Multiplicities are provided as subscripts for eigenvalues.}
{
\begin{tabular*}{\columnwidth}{@{\extracolsep{\fill}} ccc}
  \toprule
  Operators & Eigenvalues & Number of $\hat P_k$'s \\
  \midrule
  $\hat T^{0,0}_{ia}$ & $\{ 0_{6}, \pm i_{4}, \pm i2_{1} \}$  & 4 \\
  $\hat T^{0,0}_{iiaa}$ & $\{ 0_{14}, \pm i_{1} \}$  & 8 \\
  $\hat T^{0,0}_{ijaa}$, $\hat T^{0,0}_{iiab}$ & $\{0_{52}, \pm i_{4}, \pm i\sqrt{2}_{2} \}$ &  16 \\
  $\hat T^{0,0}_{ijab}$ & $\{0_{186}, \pm i_{16}, \pm i\sqrt{2}_{16}, \pm i2_{2}, \pm i2 \sqrt{2}_{1} \}$ & 32 \\
  \bottomrule
\end{tabular*} \label{tab:singlet_spectra}
}
\end{table} 
Explicit forms of the $\hat O_n$ operators for each singlet spherical operator are as follows
\bea\notag
\hat T^{0,0}_{ia} &=& \hat O_1 + \hat O_2: \\\notag
\lambda\in \{0,\pm i\}:\hat O_1 & =& \hat \kappa_{i_{\alpha}}^{a_{\alpha}} \left( \hat n_{i_{\beta}} - \hat n_{a_{\beta}} \right)^2 + \hat \kappa_{i_{\beta}}^{a_{\beta}} \left( \hat n_{i_{\alpha}} - \hat n_{a_{\alpha}} \right)^2 \\\notag
\lambda\in \{0,\pm i2\}:\hat O_2 & =& \hat T^{0,0}_{ia} - \hat O_1\\\notag
\hat T^{0,0}_{iiab} &=& \hat O_1 + \hat O_2: \\\notag
\lambda\in \{0,\pm i\}:\hat O_1 & =& \kh_{i_{\alpha} i_{\beta}}^{a_{\beta} b_{\alpha}} \left( \hat n_{a_\alpha} - \hat n_{b_{\beta}} \right)^2  \\\notag
&&+ \kh_{i_{\alpha} i_{\beta}}^{a_{\alpha} b_{\beta}} \left( \hat n_{a_{\beta}} - \hat n_{b_{\alpha}} \right)^2 \\\notag
\lambda\in \{0,\pm i\sqrt{2}\}:\hat O_2 & =& \hat T^{0,0}_{iiab} - \hat O_1\\\notag
\hat T^{0,0}_{ijaa} &=& \hat O_1 + \hat O_2: \\\notag
\lambda\in \{0,\pm i\}:\hat O_1 & =& \kh_{i_{\beta}j_{\alpha}}^{a_{\alpha} a_{\beta}} \left( \hat n_{i_\alpha} - \hat n_{j_{\beta}}  \right)^2  \\\notag 
&&+ \kh_{i_{\alpha} j_{\beta}}^{a_{\alpha} a_{\beta}} \left( \hat n_{i_{\beta}} - \hat n_{j_{\alpha}} \right)^2, \\\notag
\lambda\in \{0,\pm i\sqrt{2}\}:\hat O_2 & =& \hat T^{0,0}_{ijaa} - \hat O_1;\\\notag
\hat T^{0,0}_{ijab} &=& \sum_{i=1}^4 \hat O_i: \\ \notag
\lambda\in \{0,\pm i2\}:\hat O_1 &=&  \sum_{s, \bar{s} \in \{\alpha, \beta\}} \kh_{i_{s}j_{\bar{s}}}^{a_{\bar{s}}b_{s}} \Big( \hat n_{a_{s}} \hat n_{i_{\bar{s}}} \left( 1 - \hat n_{j_{s}} \right) \\ \notag 
+ \hat n_{b_{\bar{s}}} \hat n_{j_{s}} (1 &-& \hat n_{i_{\bar{s}}}) - \hat n_{a_{s}} \hat n_{b_{\bar{s}}} \left(\hat n_{i_{\bar{s}}} - \hat n_{j_{s}}  \right)^2 \Big), \\\notag
\lambda\in \{0,\pm i2\sqrt{2}\}:\hat O_2 &=&  \sum_{s, \bar{s} \in \{\alpha, \beta\}} \kh_{i_{s}j_{\bar{s}}}^{a_{\bar{s}}b_{s}} \Big( \hat n_{i_{\bar{s}}} \hat n_{j_{s}} \left( 1 - \hat n_{a_{s}} \right)\\ \notag
+ \hat n_{a_{s}} \hat n_{b_{\bar{s}}}  ( 1 &-& \hat n_{i_{\bar{s}}}) - \hat n_{j_{s}} \hat n_{b_{\bar{s}}} \left(\hat n_{i_{\bar{s}}} - \hat n_{a_{s}}  \right)^2 \Big), \\\notag 
\lambda\in \{0,\pm i\sqrt{2}\}:\hat O_3 &=&  \sum_{s, \bar{s} \in \{ \alpha, \beta \}} \kh_{i_{s}j_{\bar{s}}}^{a_{\bar{s}}b_{s}} \Big( \left(\hat n_{i_{\bar{s}}} - \hat n_{b_{\bar{s}}} \right)^2 \\ \notag
+ (\hat n_{j_{s}} &-& \hat n_{a_{s}} )^2   \Big) - 2 \left( \hat O_1 + \hat O_2 \right), \\\notag
\lambda\in \{0,\pm i\}:\hat O_4 &=& \hat T^{0,0}_{ijab} - \sum_{i=1}^3 \hat O_i,   
\eea
where $\hat n_p = \CR{p}\AN{p}$.
For $\Omega=0$, $\hat T^{0,0}_{iiaa}$ has only one nonzero eigenvalue and thus does not require a decomposition.

In electronic structure calculations, owing to time-reversal symmetry of the electronic Hamiltonians, unitary transformations
generating real-valued wave-functions are considered. Therefore, the technique developed in Ref.~\citenum{Kottmann:2020js} to reduce the 
number of expectation values for real fermionic wave-functions from 4 to 2 is applicable for the singlet spherical 
tensor operators as well. This leads to not more than 8 expectation values needed for evaluating gradients in the most complicated case of 
$\Omega=4$.

Note that in this case $\hat O_n$'s are not unitary operators, and thus they cannot be used by the LCU scheme. 
To estimate the number of fragments for measurements within the LCU scheme we can use  
the number of Pauli products within each fermionic rotation (see Table~\ref{tab:singlet_spectra}). 
It is clear that the number of Pauli products within singlet fermionic rotations 
can be up to 8 times larger than the corresponding number of $\hat O_n$ operators, which makes the new scheme 
much more preferable than the LCU approach.

\section{Conclusions} \label{conclusion}

We considered two approaches to generalization of the parametric-shift-rule based on the polynomial expansion 
of exponentially parametrized unitary transformations and the generator decompositions. 
As in the original parametric-shift-rule application, these approaches provide gradient 
expressions as linear combinations of expectation 
values, where the main criterion for efficiency is the number of different expectation values.  

Both of the considered approaches depend on the eigen-spectrum 
of the generator for the differentiated unitary transformation, but in different ways. 
The performance of the polynomial expansion depends only on 
the number of different eigenvalues, while that of the generator decompositions depends 
also on the generator eigen-subspaces and how well their structures can be reproduced 
by decomposing operators. The polynomial expansion approach scales quadratically with 
the number of generator eigenvalues and provides efficient expression only for 2- and 3-eigenvalue generators.
\footnote{In the 3-eigenvalue case the symmetry of the eigen-spectrum is essential.} 
For generators with a larger number of eigenvalues it is more beneficial to employ the generator decomposition technique.
{\BC This technique provides more efficient schemes (in terms of the number of needed expectation values) than any 
previous approaches for all considered generators. Also, compared to the LCU decomposition technique used 
for an arbitrary generator before,\cite{Schuld:2018gx} the new approach does not require ancilla qubits.}     

The generator decomposition approach has several variations differing in low-eigenvalue operators used for the 
decomposition. The most conservative approach is to use projectors on individual eigen-subspaces, 
its number of expectation values scales linearly with the number of the generator eigenvalues. 
It was found to be superior to other decompositions if one of the generator eigenvalues has much higher 
multiplicity than the other eigenvalues, as in the case of $S^2$-conserving fermionic operators. 

Another alternative for decomposing generators is using the Cartan sub-algebra (CSA).  
For some generators whose different eigenvalues can be related via linear combinations with binary coefficients 
and have similar degeneracies, the CSA decomposition can reduce the generator expansion to scale as $\log_2$ of the number of eigenvalues. Results of the CSA decomposition can be further improved if one will allow generation 
of non-commutative terms. 
The CSA based approaches showed that any 2-qubit transmon and match-gates require only 4 
expectation values for their gradients.   

\section{Acknowledgements}
 A.F.I. is grateful to Jakob Kottmann, Abhinav Anand, Alireza M. Khah, Lisa Jeffrey, and Ilya G. Ryabinkin 
 for stimulating discussions and acknowledges financial support from the Google Quantum Research Program,
 Early Researcher Award, and the Natural Sciences and Engineering Research Council of Canada. 
 
 \section*{Note added:}
After submission of this manuscript to arXiv, two more proposals generalizing the PSR via methods identical to our 
polynomial expansion were submitted.\cite{wierichs2021general,kyriienko2021generalized} 
 
 \section{Appendix A}
 
 Here we derive \eqs{eq:exp} and \eqref{eq:poly} for 
 generator $\hat G$ that has $L$ eigenvalues. First, to find $a_i(\theta)$'s in \eq{eq:poly} we use 
 an eigen-space projector decomposition of $\hat G$:
 \bea
 \hat G = \sum_{n=1}^L \hat P_n \lambda_n,
 \eea
 where $\lambda_n$ are different eigenvalues of $\hat G$ and $\hat P_n$ are projectors on the corresponding 
 eigen-subspaces. Convenient properties of these projectors are their orthogonality and idempotency ($\hat P_n\hat P_m=\delta_{nm}\hat P_n$). These properties allows us to connect the exponential function
 \bea  
e^{i\theta \hat G} = \sum_{n=1}^L \hat P_n e^{i\theta\lambda_n},
\eea
with its polynomial expansion
 \bea  
\sum_{k=0}^{L-1} a_k(\theta) (i\hat G)^k= \sum_{n=1}^L \hat P_n \sum_{k=0}^{L-1} a_k(\theta) (i\lambda_n)^k.
\eea
Due to linear independence of projector operators this results in the system of linear equations with $\{i^k a_k(\theta)\}$ as variables
\bea
 \sum_{k=0}^{L-1} \lambda_n^k [i^k a_k(\theta)] = e^{i\theta\lambda_n},\quad n=1,...,L.
\eea
The matrix involved in this system of equations is the Vandermonde matrix ($\lambda_n^k = W_{nk}$), 
whose determinant is non-zero as long as the eigenvalues are different. 
Inverting the Vandermonde matrix provide $a_k(\theta)$ solutions 
\bea
a_k(\theta) = i^{-k}\sum_n W^{-1}_{kn} e^{i\theta\lambda_n}.
\eea
Since $\lambda_n$'s are real, it is easy to show the following relations 
\bea
a_{2k}(\theta) &=& a_{2k}^*(-\theta), \\
a_{2k+1}(\theta) &=& -a_{2k+1}^*(-\theta).  
\eea

Second, $C_n$ coefficients in \eq{eq:exp} can be found as solutions of a linear system of equations.
This system can be formulated by rewriting \eq{eq:exp} as  
\bea\notag
\sum_n C_n \langle e^{-i\theta_n \hat G} \tilde{H} e^{i\theta_n \hat G} \rangle &=& \sum_{n} C_n \sum_{k,k'=0}^{L-1} \langle \hat G^{k'}  \tilde{H} \hat G^k \rangle A_{kk'}(\theta_n) \\ 
&=& i\left[ \langle \tilde{H} \hat G \rangle - \langle \hat G \tilde{H} \rangle\right],
\eea
where 
\bea
\tilde{H} &=& e^{-i\tau \hat G} \hat U_2^\dagger \hat H \hat U_2 e^{i\tau \hat G},\\
A_{kk'}(\theta_n) &=& a_k(\theta_n) a_{k'}^*(\theta_n) i^{k+k'} (-1)^k.
\eea
Accounting for linear independence of $\langle \hat G^{k'}  \tilde{H} \hat G^k \rangle$ terms one can 
obtain $C_n$ from equations 
\bea
\sum_n A_{kk'}(\theta_n) C_n = B_{kk'},
\eea
where $B_{10}= -B_{01} = i$ and $B_{kk'} = 0$ for all other $kk'$. 
Depending on the choice of $\theta_n$, the number of needed $C_n$ to satisfy $L^2$ equations 
can vary, but it cannot exceed $L^2$. 

Minimization of the number of $C_n$ coefficients and thus the number of expectation values depends 
on the $\hat G$ spectrum. For example, if every $\lambda_n$ has its negative counterpart, $-\lambda_n$,
then the even (odd) degree functions $a_{2k}(\theta)$ ($a_{2k+1}(\theta)$) are real even (odd) $\theta$ functions.
This condition allows one to reduce the number of needed $C_n$ and $\theta_n$ parameters to $\sim L^2/2$, where 
$\theta_n$'s are chosen in pairs $\{\pm \theta_{k}\}_{k=1}^{N_p/2}$. Thus, 
in the case of $L=2$, the number of $C_n$'s is only 2 because $\theta_n=\pm \theta$ 
creates some dependencies in $A_{kk'}(\theta_n)$ elements. 

For $L=3$ and $\lambda_n \in \{0,\pm 1\}$ 
one can derive the following polynomial expansion of the exponential $\hat G$ operator
\bea
e^{i\theta \hat G} = 1 + i\sin(\theta)\hat G+(\cos(\theta)-1)\hat G^2.
\eea
Taking $\theta_{1,2} = \pm \theta$ does not eliminate terms $\langle \hat G \tilde{H} \hat G^2\rangle$ 
and  $\langle \hat G^2 \tilde{H} \hat G\rangle$ in the PSR expression, therefore another pair of 
$\theta$'s $\theta_{3,4} = \pm 2\theta$ are needed to eliminate these terms and to obtain the 
gradient of energy in this case. Here, we present the final expression
\bea
 i \langle [\tilde{H}, \hat G] \rangle  = (\alpha\Delta_1-\Delta_2)\beta,
\eea
where 
\bea
\alpha &=& \frac{\sin(2\theta)(\cos(2\theta)-1)}{\sin(\theta)(\cos(\theta)-1)}, \\
\beta &=& \frac{1}{2\sin(2\theta)}\left[\frac{1-\cos(2\theta)}{1-\cos(\theta)} -1\right]^{-1}, \\
\Delta_k &=& \langle e^{-ik\theta \hat G} \tilde{H} e^{ik\theta \hat G} \rangle - \langle e^{ik\theta \hat G} \tilde{H} e^{-ik\theta \hat G} \rangle.
\eea
This results in 4 expectation values required to obtain the gradient with respect to the amplitude of 
the $L=3$ $\hat G$ with the symmetric eigenvalue spectrum $\lambda_n \in \{0,\pm 1\}$. 


%

\end{document}